\documentstyle[12pt]{article}
\begin{document}
%
%
\newcommand{\bl}{\begin{flushleft}}
\newcommand{\br}{\begin{flushright}}
\newcommand{\el}{\end{flushleft}}
\newcommand{\er}{\end{flushright}}
\newcommand{\bc}{\begin{center}}
\newcommand{\ec}{\end{center}}
\newcommand{\del}{\partial}
\newcommand{\ba}{\begin{array}}
\newcommand{\ea}{\end{array}}
\newcommand{\eq}[2]{\begin{equation}#2\label{#1}\end{equation}}
\newcommand{\eqn}[2]{\begin{eqnarray}#2\label{#1}\end{eqnarray}}
\newcommand{\nn}{\nonumber\\}
\newcommand{\grad}{\nabla}
\newcommand{\vev}[1]{\langle #1\rangle}
\newcommand{\jour}[4]{\ \em{#1}\/\ \bf{#2},\ \rm{#3\ (#4)}}
\newcommand{\book}[3]{\ \em{#1},\/\ \rm{#2\ (#3)}}
\newcommand{\listrefs}[1]{\begin{thebibliography}{999}#1
                           \end{thebibliography}}
\newcommand{\prepno}[2]{\hfill{\vbox{\hbox{#1}\hbox{#2}}}}
\newcommand{\Title}[1]{\begin{center}\huge{\bf #1}\end{center}
                                      \vspace{.5cm}}
\newcommand{\Author}[3]{\begin{center}\Large{\bf #1}\footnote{
              electronic address: #2}\\
  \smallskip\normalsize{\em #3}\\ \medskip\today\end{center}
                                   \vspace{.5cm}}
\newcommand{\Authors}[5]{\begin{center}\Large{\bf #1}\footnote{
              electronic address: #3} and 
             {\bf #2}\footnote{electronic address: #4}\\
   \smallskip\normalsize{\em #5}\\ \medskip\today\end{center}
                                    \vspace{.2cm}}
\newcommand{\Abstract}[1]{\begin{center}\Large{Abstract}\\
     \medskip\begin{quote}\small{#1}\end{quote}\end{center}}
\newcommand{\listindex}[1]{\begin{theindex}#1\end{theindex}}
\newcommand{\subs}[1]{\subsection{#1}\setcounter{equation}{0}}
\renewcommand{\theequation}{\thesection.\arabic{equation}}
\newcommand{\newsec}[1]{\section{#1}\setcounter{equation}{0}}
\newcommand{\bb}{\begin{eqnarray}}
\newcommand{\ee}{\end{eqnarray}}
%
%
\newcommand{\half}{{1\over 2}}
\newcommand{\laplace}{{\kern1pt\vbox{\hrule height 1.2pt\hbox
            {\vrule width 1.2pt\hskip 3pt\vbox{\vskip 6pt}\hskip 3pt\vrule 
        width 0.6pt}\hrule height 0.6pt}\kern1pt}}
\renewcommand{\slash}[1]{#1\!\!\!\! /}
\newcommand{\wt}{\widetilde}
\newcommand{\Z}{{\cal Z}}
\newcommand{\D}{{\cal D}}
\newcommand{\mup}{\mu^{\prime}}
\newcommand{\nup}{\nu^{\prime}}
\newcommand{\intg}{\int d^2x\sqrt{-g}}
\newcommand{\psid}{\psi_{\delta}}
%
\newcommand{\PR}{Phys. Rev.}
\newcommand{\PRL}{Phys. Rev. Lett.}
\newcommand{\RMP}{Rev. Mod. Phys.}
\newcommand{\NP}{Nucl. Phys.}
\newcommand{\PL}{Phys. Lett.}
\newcommand{\PREP}{Phys. Rep.}
\newcommand{\CMP}{Comm. Math. Phys.}
\newcommand{\JMP}{J. Math. Phys.}
\newcommand{\CGG}{Class. Quant. Grav.}
\newcommand{\MPL}{Mod. Phys. Lett.}
\newcommand{\IJMP}{Int. J. Mod. Phys.}
\newcommand{\AP}{Ann. Phys.}
%
%
%
%
\begin{titlepage}
\prepno{SINP/TNP/96-06}{hep-th/9604056}
\Title{QED$_2$ in Curved Backgrounds}
\Author{Amit Ghosh}{amit@tnp.saha.ernet.in}
        {Saha Institute of Nuclear Physics,\\ 1/AF 
                     Bidhannagar, Calcutta 700 064, INDIA}
\Abstract{Here we discuss the two dimensional
          quantum electrodynamics in curved space-time, especially in the 
          background of some black holes. We first show
          the existence of some new quantum mechanical solution
          which has interesting properties. Then for some special black holes 
          we discuss the fermion-black hole scattering problem. The issue
          of confinement is intimately connected with these solutions and
          we also comment on this in this background. Finally, the 
          entanglement entropy and the Hawking radiation are also discussed 
          in this background from a slightly different viewpoint.}
\end{titlepage}
%
%
\newsec{Introduction}
The  Schwinger  model  \cite{SCH}, in curved  two-dimensional
space-time, has been studied over the  years \cite{miscl} and it has been
speculated that the qualitative behavior of the model should 
not change in the presence of gravity \cite{curved}. The comparison
can be made with the finite temperature QED$_2$ \cite{temp} where
no non-trivial phase appears and the model stays only in one phase
which is the screened Coloumb phase that exists at the zero
temperature. However, a rigorous  evidence about this was missing
for the curved backgrounds. In this paper we develop a general
formalism of the model for an arbitrary curved background and finally
gather evidences about the phases for some particular background. What
we observe is that the phase structure remains unaltered at least for
this special background and the model may exist in two phases, namely
in screened Coulomb and the unconfining phases, just as it
was in the flat case. This observation has crucial impact on the study
of fermion-black hole scattering problem which was done in an earlier paper
\cite{GM}. Earlier we neglected the gravitational interactions and found that
the model itself can not avoid the problem of information loss. In
this paper we shall take up the same problem but now not neglecting
the gravity. However, as the model doesn't show any qualitative change
the conclusions are obvious - the model still supports the information 
loss. The important lesson that we learn from this is that this
problem can possibly be avoided only if we incorporate the quantum
gravitational effects.

In recent years another important area of interest has been the study
of matter fields in the black hole backgrounds. It provides a lot of
insights into the problem of black hole entropy, the study of Hawking
radiation and all these raised more issues about the quantum theory
of gravity. As the spectrum of QED$_2$ contains only a scalar field
the study of QED$_2$ in black hole backgrounds boils down to the
study of scalar fields. In this paper we study this problem, namely
we first calculate the entanglement entropy of a scalar field in the
particular black hole background which we considered in section 1 and
discuss the Hawking radiation of this black hole. We show that the whole
analysis can be done with sufficient simplicity.

\newsec{QED$_2$ in curved background}
QED$_2$ or the Schwinger model in curved space-time is described by the 
Lagrangian density \cite{SCH}
\eq{31}{{\cal L}=-{1\over 4}\,g^{\mu\mup}g^{\nu\nup}F_{\mu\nu}
       F_{\mup\nup}+i\,\overline\psi\gamma^{\mu}(x)D_{\mu}\psi.}
where the indices $\mu,\nu...$\,etc. refer to the curved background and take  
the values 0,1. Other notations are standard. To introduce the fermions we
need to go to a locally flat space-time with which the correspondences are
established via the zweibeins. The zweibeins satisfy the relations
\eqn{32}{e^{\mu a}(x)\gamma_a &=&\gamma^{\mu}(x),\quad e^{\mu a}(x)
          e^b_{\mu}(x)=
        \eta^{ab},\nn e^{\mu a}(x)e_{\nu a}(x)&=&\delta^{\mu}_{\nu},\qquad 
        e^{\mu a}(x)e_{\mu b}(x)=\delta^a_b.}
The indices $a,b...$etc. are denoting the flat space-time, the flat
space indices are raised and lowered by the metric $\eta_{ab}$ and
the curved space indices are raised and lowered by $g_{\mu\nu}$.
The field strength being antisymmetric in the space-time indices continues
to have the same form $F_{\mu\nu}=\del_{\mu}A_{\nu}-\del_{\nu}A_{\mu}$.
However, the gauge covariant derivative acting on the fermions is of the form,
$D_{\mu}\psi=(\grad_{\mu}-ieA_{\mu})\psi$, where $\grad_{\mu}=\del_{\mu}+
\half\,\omega_{\mu}^{ab}\sigma_{ab}$. $\sigma_{ab}={1\over 4}[\gamma_a,\gamma_b]$
are the standard Lorentz transformation generators and $\omega_{\mu}^{ab}$ 
are the spin connections. We shall be working in the gauge $\grad_{\mu}
e^{\nu a}=0$, which fixes the spin connections completely in terms of the
zweibeins $\omega_{\mu ab}=\half\,[e^{\nu}_a(\del_{\mu}e_{\nu b}-\del_{\nu}
e_{\mu b})+\half e^{\rho}_ae^{\sigma}_b(\del_{\sigma}e_{\rho c})e^c_{\mu}-
(a\leftrightarrow b)]$. Throughout we shall be using the following notations
and conventions: for flat space $\eta^{ab}={\rm diag}(1,-1),\gamma^5=\gamma^0
\gamma^1,(\gamma^0)^2=-(\gamma^1)^2=1$ and $\epsilon_{01}=+1$. For the curved
background $\sqrt{-g}\gamma^{\mu}\epsilon_{\mu\nu}=\gamma^5\gamma_{\nu}$ and
$\wt\grad_{\mu}=\sqrt{-g}\epsilon_{\mu\nu}\grad^{\nu}$.

The partition function is given by
\eq{33}{\Z=\int\D A_{\mu}\D\overline\psi\,\D\psi\, e^{\,iS},\qquad S=\intg\,{\cal L}.}
The effective action is defined by the following functional of the abelian 
gauge field $A_{\mu}$
\eq{34}{e^{i\Gamma[A]}=\int{\cal D}\overline\psi\,{\cal D}\psi\, e^{i\intg
       \;\overline\psi i\slash D\,\psi}.}
Now in two dimensions we can always set
\eq{35}{A_{\mu}=-{\sqrt\pi\over e}\,(\wt\grad_{\mu}\sigma+\grad_{\mu}\wt\eta\,)}
where, $\sigma$ and $\wt\eta$ are scalar fields. So the field strength is 
given by $F_{\mu\nu}={\sqrt\pi\over e}\epsilon_{\mu\nu}\sqrt{-g}\,\laplace
\sigma$ where, $\laplace\sigma={1\over\sqrt{-g}}\,\del_{\mu}(g^{\mu\nu}\sqrt{
-g}\del_{\nu}\sigma)$. The Dirac operator is given by
\eq{36}{\slash D=\slash\grad+i\sqrt\pi\slash\grad\,\wt\eta+i\sqrt\pi\gamma^5
       \slash\grad\,\sigma.}

It is easy to see that the transformations   
\eqn{37}{\psi &\to& e^{\,i\sqrt\pi\,(\widetilde\eta-\gamma_5\sigma)}\,\psi,\nn 
        \overline{\psi}&\to& \overline{\psi}\,e^{-i\sqrt{\pi}\,(\wt\eta+\gamma_5
        \sigma)},}
decouple  the  gauge field from the fermions and the classical
action becomes free, {\it i.e.},
\eq{38}{\overline{\psi}i\slash D\,\psi\to\overline\psi i\slash\partial\,\psi.}
However, we should proceed through infinitesimal steps. For the 
time being it is sufficient to consider only the chiral transformations since
the effective action is known to be invariant under the standard gauge
transformations. But quantum mechanically the chiral symmetry becomes 
anomalous because the fermionic measure does not remain invariant under those
transformations. To see this explicitly let us make the following infinitesimal
chiral redefinition of the fermionic variables,
\eqn{39}{\psi &\to&\psid=(1-i\sqrt\pi\gamma^5\delta\sigma)\,\psi,\nn
        \overline\psi &\to&\overline\psid=\overline\psi\,(1-i\sqrt\pi
        \gamma^5\delta\sigma)}
leading to
\eqn{310}{\overline\psi\slash D\,\psi &=&\overline\psid\,[\slash\grad+i\sqrt\pi
        \gamma^5\slash\grad\,(\sigma-\delta\sigma)]\,\psid\nn&=&\overline\psid
        \slash D\,\psid-i\sqrt\pi\,\overline\psid\gamma^5\slash\grad\,(\delta
        \sigma)\,\psid.}
So the effective action becomes
\eqn{311}{\Z &=&\int\D\overline\psi\,\D\psi\, e^{\,iS}\nn
            &=&\int\D\overline\psid\,\D\psid\, e^{\,iS_{\delta}}\nn
            &=&\int\D\overline\psi\,\D\psi\, e^{\,iS}\,[1+i\sqrt\pi\intg\,\overline\psi\gamma^5\slash\grad\,(\delta\sigma)\psi]\nn
            &=&\Z+i\sqrt\pi\int\D\overline\psi\,\D\psi\, e^{\,iS}\intg\,\delta\sigma
                \grad_{\mu}J^{\mu}_5.}
where, $J^{\mu}_5=\overline\psi\gamma^{\mu}\gamma^5\psi$. This apparently
shows that $\grad_{\mu}J^{\mu}_5=0$ which in turn implies the presence of
chiral invariance. However, it is well known that $\D\overline\psid\D\psid
\neq\D\overline\psi\,\D\psi$. To calculate explicitly the Jacobian we proceed
as follows.

We first analytically continue the space-time to the Euclidean domain, i.e. 
$x^0\to -ix^4,\,\gamma^0\to i\gamma^4,\,D^0\to iD^4$. Then we choose a set
of complete orthonormal functions $\{\Phi_n(x)\}$ satisfying
\eq{312}{\int d^2x\sqrt g\,\Phi_n^{\dagger}\Phi_m=\delta_{nm}\qquad\sum_n
        \sqrt g\,\Phi_n(x)\Phi^{\dagger}_n(y)=\delta^2(x-y).} 
The fermionic fields can be expanded in terms of these functions as
$\psi(x)=\sum_na_n\Phi_n(x)$ and $\overline\psi(x)=\sum_n\Phi_n^{\dagger}
(x)\,b_n$, where $a_n$ and $b_n$'s are Grassmann numbers. The fermionic 
measures are accordingly expressed as $\D\psi=\prod_n
da_n$ and $\D\overline\psi=\prod_ndb_n$. Therefore, under the chiral
transformation $\psi\to\psid=\sum_na_n^{\delta}\Phi_n\,=\sum_{nm}C_{nm}a_m
\Phi_n(x)$ the measure changes as $\D\psid=\prod da_n^{\delta}={1\over \det C}
\prod_nda_n={1\over \det C}\D\psi$. So the Jacobian is the determinant of
the matrix 
\eq{312a}{C_{nm}=\delta_{nm}-i\sqrt\pi\int d^2x\sqrt g\,\delta\sigma(x)
         \Phi_n^{\dagger}(x)\gamma^5\Phi_m(x).} 
So
\eqn{313}{\det C&=&\exp\,({\rm Tr}\ln C)\nn 
              &=&\exp\,[-i\sqrt\pi\int d^2x\sqrt g\,\delta\sigma(x)\sum_n
            \Phi_n^{\dagger}(x)\gamma^5\Phi_n(x)].}
The entire measure is just the square of this
\eq{314}{\D\overline\psid\D\psid={1\over (\det C)^2}\,\D\overline\psi\D\psi
        =\D\overline\psi\D\psi\,\exp\,\Big[2i\sqrt\pi\int d^2x\sqrt g\,\delta\sigma(x)
          \sum_n\Phi_n^{\dagger}(x)\gamma^5\Phi_n(x)\Big].}
The sum in the exponent is divergent since arbitrary higher modes go into 
the summation. We shall adopt here the regularization procedure demonstrated
by Fujikawa \cite{FUJI} by letting the higher modes to damp exponentially. 
For this let
us consider a gauge invariant Dirac like operator whose eigenfunctions are to
be identified with these complete set of orthonormal functions. The operator
must be chosen to be gauge invariant in order to ensure that the gauge 
invariance of the measure is not destroyed in the process of regularization.
The operator also need to be Dirac like since otherwise the trace over the
Dirac gamma matrices will produce a trivial zero result. So in order to avoid
the triviality and restore the gauge invariance one is sufficiently 
restricted in the choice of the regularizing operator. One natural choice,
which Fujikawa himself employed \cite{FUJI} has been the Euclidean Dirac 
operator of the
action itself. However, after that many people \cite{HAG,FJ} realized that 
this is not that
sacred and one such explicit varied choice was suggested in \cite{BGM}. It 
seems to be rather unique general possibility obeying the above constraints
if one combines with them the requirement that the linearity of the theory is
to be maintained. As the theory in a curved background doesn't have a global
translation invariance the regularizing operator can more generally be taken
similar to \cite{GM}
\eq{315}{\slash D_r\Phi_n=\lambda\Phi_n,\qquad\slash D_r=\gamma_{\mu}D^r_{\mu}
        =\gamma_{\mu}(\grad_{\mu}-ieA^r_{\mu})}
where $A_{\mu}^r$ is the regularizing background which is taken to be 
$A^r_{\mu}=A_{\mu}-\grad_{\nu}(\varphi(x)F_{\mu\nu})$. The scalar field 
$\varphi$ is breaking the global translation invariance. We have taken $\varphi$ 
to be a function of the space only.

Now usually in flat space-time we use the plane wave representation of the
basis $\{\Phi_n\}$ to evaluate the trace. But in the curved background due to
the loss of global translation invariance there is no global representation of 
the momentum. But this can be achieved in the Riemann normal coordinates 
$\xi^{\mu}=(x-y)^{\mu}$ and
the use of these coordinates is reasonable since we will be working in the 
small space-time regions to calculate the trace. So we proceed as follows
\eqn{316}{\left[\int d^2x\sqrt g\,\delta\sigma(x)\,\Phi_n^{\dagger}\gamma^5\Phi_n
         \right]_{\rm reg}&=&\lim_{M^2\to\infty}\sum_n\int d^2x\sqrt g\,\delta
         \sigma(x)\,\Phi_n^{\dagger}(x)\gamma^5\Phi_n(x)e^{-\lambda_n^2/M^2}\nn
         &=&\lim_{M^2\to\infty}\sum_n\int d^2x\sqrt g\,\delta\sigma(x)\Phi_n^{
             \dagger}(x)\gamma^5e^{-\slash D_r^{2}/M^2}\Phi_n(x)\nn
         &=&\lim_{M^2\to\infty\atop x\to y}\int d^2x\,\delta\sigma(x)\;{\rm Tr}
             \,\gamma^5e^{-\slash D_r^{2}/M^2}\delta^2(x-y)\nn 
         &=&\lim_{M^2\to\infty\atop\xi\to 0}\int d^2x\delta\sigma(x)\int{d^2p
             \over (2\pi)^2}{\rm Tr}\gamma^5e^{-\slash D_r^{2}/M^2}
              e^{-ip\cdot\xi}}
It is necessary to calculate the explicit form of the operator $\slash D_r^2$
and then the trace of the gamma matrices and the limits are to be taken
carefully. First using the formula $R_{\mu\nu}^{\,ab}=\del_{\mu}\omega_{\nu}^{
ab}+\omega_{\mu}^{ac}\omega_{\nu}^{cb}-(\mu\leftrightarrow\nu)$ we get $\slash
D_r^{\,2}=D^r_{\mu}D^r_{\mu}+{1\over 12}\,R+{e\over 2\sqrt g}\,\epsilon_{\mu\nu}
\gamma^5F_{\mu\nu}^r$, where $F^r_{\mu\nu}=\del_{\mu}A^r_{\nu}-\del_{\nu}A^r_{
\mu}$. Using these
\eq{317}{\left[\int d^2x\sqrt g\,\delta\sigma(x)\Phi_n^{\dagger}\gamma^5\Phi_n
         \right]_{\rm reg}=-{e\over 4\pi}\int d^2x\,\delta\sigma(x)\epsilon_{\mu
          \nu}F^r_{\mu\nu}.}
So, the Jacobian is 
\eq{318}{\D\overline\psid\D\psid=\D\overline\psi\D\psi\left[1-{ie\sqrt\pi\over 
         2\pi}\int d^2x\,\delta\sigma(x)\epsilon_{\mu\nu}F^r_{\mu\nu}\right].}
Thus there is a chiral anomaly given by
\eq{319}{\grad_{\mu}J^{\mu}_5={1\over\sqrt{-g}}\,{e\over 2\pi}\,\epsilon^{\mu\nu}
        F^r_{\mu\nu}.}
Using the expression ${e\over 2\pi}\,\epsilon^{\mu\nu}F^r_{\mu\nu}=-\sqrt{-g}
{1\over\sqrt\pi}\,\laplace\,(\sigma+\varphi\laplace\sigma)$
\eq{320}{\grad_{\mu}J^{\mu}_5=-{1\over\sqrt\pi}\,\laplace\,(\sigma+\varphi
         \laplace\sigma).}
Let us now go to calculate the effective action/bosonized action. By a small
change of variables, both in $\sigma$ and $\wt\eta$, we find that the measure
is offering a Jacobian only under the change of the chiral variables. Thus we
arrived at the equation
\eq{321}{e^{\,i\Gamma[\,\sigma,\,\wt\eta\,]}=(\det C_{\rm reg})^2\,e^{\,i\Gamma[\,\sigma-
        \delta\sigma,\,\wt\eta-\delta\wt\eta\,]}.}
Making the Taylor expansion of the effective functional we get
\eq{322}{\delta\Gamma[\sigma]=\intg\,\delta\sigma(x)\laplace\,(\sigma+\varphi
         \laplace\sigma),\qquad{\delta\Gamma\over\delta\wt\eta}=0}
leading to
\eq{323}{\Gamma[\sigma]=\half\intg\,\left[\sigma\laplace\sigma+\varphi\laplace
         \sigma\laplace\sigma\right].} 
Re-expressing this in terms of the gauge fields 
\eq{324}{\Gamma[A]=\intg\left[\,{e^2\over 2\pi}\,\wt\grad\cdot A\,{1\over\laplace}\,\wt
         \grad\cdot A-{e^2\varphi\over 4\pi}\,g^{\mu\mup}g^{\nu\nup}F_{\mu\nu}
         F_{\mup\nup}\right].}
The local form of the resulting bosonized action can be obtained by 
introducing an auxiliary field $\Sigma$
\eq{325}{S_B=\intg\left[-{1\over 4}(1+{e^2\varphi\over\pi})\,g^{\mu\mup}g^{\nu
         \nup}F_{\mu\nu}F_{\mup\nup}+{e^2\over 2\pi}A^2+\half\,\grad_{\mu}\Sigma
        \grad^{\mu}\Sigma-{e\over\sqrt\pi}A^{\mu}\grad_{\mu}\Sigma\right].}
Then one can do the standard constraint analysis to calculate the Hamiltonian.
First, the canonical momenta have to be defined. The momenta corresponding to 
$A_0, A_1$ and $\Sigma$ are respectively
\eqn{326}{\Pi^0&=&{\delta S_B\over\delta\grad_0A_0}=0\nn
         \Pi^1&=&{\delta S_B\over\delta\grad_0A_1}
               ={1\over\sqrt{-g}}\Bigl(1+{e^2\varphi\over\pi}\Bigr)
                 (\grad_0A_1-\grad_1A_0)\nn
         \Pi^\Sigma &=&{\delta S_B\over\delta\grad_0\Sigma} 
                     =\sqrt{-g}\,(\grad^0\Sigma-{e\over\sqrt\pi}A^0).}
The first of these equations is recognized to be a constraint. Using all 
these equations, we obtain the Hamiltonian
\eqn{327}{{\cal H}&=&\Pi^1\grad_0A_1+\Pi^{\Sigma}\grad_0\Sigma-{\cal L}_B\nn
          &=&{\sqrt{-g}\,(\Pi^1)^2\over 2\,(1+ e^2\varphi/\pi)}+\Pi^1\grad_1
          A_0-\half\sqrt{-g}\,g^{11}(\grad_1\Sigma)^2+{e\over\sqrt\pi}\sqrt{-g}
          \,A^1\grad_1\Sigma-{e^2\over 2\pi}\sqrt{-g}\,A^2\nn &&+\,{1
          \over 2g^{00}\sqrt{-g}}\left[\,\Pi^{\Sigma}+\sqrt{-g}
          ({e\over\sqrt\pi}A^0-g^{01}\grad_1\Sigma)\right]^2.}
The consistency that the first constraint equation be invariant under time 
evolution by this Hamiltonian requires a secondary constraint which is the
Gauss law
\eq{328}{G\equiv\grad_1\Pi^1-{e\over\sqrt\pi}\,\Pi^\Sigma=0.}
There are no further constraints, and it can be checked that the Poisson 
brackets of $G$ and $\Pi^0$ vanish, so that the 
constraints are {\it first class}. This is natural, as we have taken care 
to maintain gauge invariance in the effective action. As usual, then, we 
have to fix a gauge to remove gauge degrees of freedom. It is convenient 
here to consider the physical gauge conditions
\eq{329}{\Sigma=A_0=0.}
In the present gauge, the Hamiltonian simplifies to 
\eq{330}{{\cal H}={\sqrt{-g}\,(\Pi^1)^2\over 2\,(1+ e^2\varphi/\pi)}+{1\over
        g^{00}}\grad_1\Pi^1g^{01}A_1+{e^2\over 2\pi}{A_1^2\over g^{00}
        \sqrt{-g}}+{\pi\over e^2}{(\grad_1\Pi^1)^2\over 2\,g^{00}\sqrt{-g}}.}
Note that ${\cal H}$ is preserved in time since we allow $\varphi$ to depend
on space only.
Now it is interesting to notice that this can be brought to a Hamiltonian
of a ``free field" in a curved background. We should keep in in mind that
a ``free field" Hamiltonian in a gravitational field is no longer really a
free theory as the particles can interact gravitationally. If we start from
the Lagrangian of a scalar field in a curved space-time  
\eq{331}{{\cal L}=\half\sqrt{-g}\,(\grad^{\mu}\Phi\grad_{\mu}\Phi-M^2\Phi^2),}
the Hamiltonian takes the form
\eq{332}{{\cal H}={1\over 2g^{00}\sqrt{-g}}\,(\Pi^{\Phi}-\sqrt{-g}g^{01}\grad_1
         \Phi)^2-\half\sqrt{-g}\,g^{11}(\grad_1\Phi)^2+\half\sqrt{-g}\,M^2\Phi^2}
where $\Pi^{\Phi}=\sqrt{-g}\,\grad^0\Phi$ is the canonically conjugate momentum
of $\Phi$. Comparing (\ref{330}) and (\ref{332}) our Hamiltonian may be 
converted to the familiar ``free field" form by the redefinitions
\eq{333}{\Phi={\sqrt\pi\over e}\,\Pi^1,\qquad\Pi^\Phi=-{e\over\sqrt\pi}A_1.}
This  shows that the physical spectrum of the model contains just a massive 
boson with mass $M=e/\sqrt{\pi+e^2\varphi}$. 

Let us investigate the nature of the force mediated by the  gauge
field of this theory between two quarks. First,
in   the   presence   of   two  static external quarks
($q\overline q$-pair) of
charge $Q$ at $\pm L/2$, the charge density is modified to
\begin{eqnarray}
J_Q^0(t,x^1)&=&{Q\over e}\sqrt{g_{00}\over g}\,\Bigl[\,\delta(x^1-{L\over 2})
 -\delta(x^1 +{L\over  2})\,\Bigr]+J^0\nonumber\\
&=&J^0-{1\over{\sqrt\pi}}\,\grad_1\chi,\end{eqnarray}
where,
\begin{equation}
\chi={Q\sqrt\pi\over e}\,\theta\,(x^1+{L\over 2})\,\theta\,({L\over
2}-x^1)\sqrt{g_{00}\over g}.
\end{equation}  Remembering that $eJ_\mu=\delta\Gamma[A]/\delta A^\mu$ 
the  Lagrangian  density in the presence of these external quarks can be 
written as 
\begin{equation}
{\cal L_Q}={\cal L}_B-{e\over{\sqrt\pi}}\sqrt{-g}\,\widetilde{\grad}\cdot A
\,\chi.  \end{equation}
The momenta corresponding to $\chi,A_0, A_1$ and $\Sigma$ are respectively
\eqn{337}{\Pi^\chi_Q &=&{\del{\cal L_Q}\over\del\grad_0\chi}=0\nn
         \Pi^0_Q &=&{\del{\cal L_Q}\over\del\grad_0A_0}=0\nn
         \Pi^1_Q &=&{\del{\cal L_Q}\over\del\grad_0A_1}
               =\Pi^1-{e\over\sqrt\pi}\,\chi\nn
         \Pi^\Sigma_Q &=&{\del{\cal L_Q}\over\del\grad_0\Sigma}=\Pi^\Sigma.}
The first two of these equations are recognized to be primary constraints. 
Using all these equations, we obtain the Hamiltonian
\eqn{338}{{\cal H_Q}&=&\Pi^1_Q\grad_0A_1+\Pi^\Sigma_Q\,\grad_0\Sigma
-{\cal L_Q}\nn &=&{\cal H}-{e\over\sqrt\pi}\grad_1A_0\,\chi.}
The consistency that the primary constraint equations be invariant under time 
evolution by this Hamiltonian requires secondary constraints which are 
\eqn{339}{G_1&\equiv &\grad_1\Pi^1_Q-{e\over\sqrt\pi}\,\Pi^\Sigma=0\nn
       G_2&\equiv &{e\over\sqrt\pi}\grad_1A_0=0.}
There are no further constraints, and it can be checked that the mutual 
Poisson brackets of the constraints  with one another vanish, so that the 
constraints are {\it first class}. This is natural, as we have taken care 
to maintain gauge invariance in the effective action. As usual, then, we 
have to fix a gauge to remove gauge degrees of freedom. It is convenient 
here to consider the physical gauge conditions
\eq{340}{\Sigma=A_0=0.}
In the present gauge, the Hamiltonian simplifies to 
\eq{341}{{\cal H_Q}={\sqrt{-g}\,(\Pi^1)^2\over 2\,(1+ e^2\varphi/\pi)}+{1\over
        g^{00}}\grad_1\Pi^1_Q\,g^{01}A_1+{e^2\over 2\pi}{A_1^2\over g^{00}
        \sqrt{-g}}+{\pi\over e^2}{(\grad_1\Pi^1_Q)^2\over 2\,g^{00}\sqrt{-g}}.}
Now it is interesting to notice that this can also be brought to a Hamiltonian
almost similar to a ``free one'' by the following redefinition of fields
\eq{342}{\wt\Pi^\Phi=\Pi^\Phi,\qquad\wt\Phi=\Phi+\chi={\sqrt\pi\over e}\,\Pi^1_Q}
leading to
\eq{343}{{\cal H_Q}={(\wt\Pi^\Phi)^2\over 2\,g^{00}\sqrt{-g}}-{1\over g^{00}}
       \,\wt\Pi^{\Phi}g^{01}\grad_1\wt\Phi^2+{(\grad_1\wt\Phi)^2\over 2g^{00}
       \sqrt{-g}}+\half\sqrt{-g}{e^2\over\pi+e^2\varphi}\,(\wt\Phi-\chi)^2}
Then the potential between the quark-antiquark pair would be the difference
between the ground state energies of these two Hamiltonians ${\cal H_Q}$
and ${\cal H}$ and this can be calculated because both the Hamiltonians
are still quadratic in the momenta. The straightforward path-integral
evaluation gives
\begin{equation}
V(L)=E_Q-E={1\over 2}\int dx^1\sqrt{-g}\left[{e^2\over\pi+e^2\varphi}\chi^2 
+({e^2\over\pi+e^2\varphi}\chi){1\over\grad^1\grad_1-{e^2\over\pi+e^2\varphi}}
(\chi{e^2\over\pi+e^2\varphi})\right].
\end{equation}
If we compare this expression with the potential we obtain in the flat
background we see that there is great similarity between them. For our
purpose let us take the expression given in \cite{BGM}
\eq{344a}{V(L)={1\over 2}\int dx^1\left[\,{e^2\over\pi+e^2a}\,\chi^2 
          +({e^2\over\pi+e^2a}\,\chi)\,{1\over\del_1^2-{e^2\over\pi+e^2a}}
          \,(\chi{e^2\over\pi+e^2a})\,\right]}
where in $\chi$ here we have to put the flat Minkowski metric and also
$\varphi=a$ is a constant parameter.
It is almost impossible to calculate the nature of the potential for an 
arbitrary background. So we consider some special case. In the next section we
shall be considering  a particular example of a $(1+1)$-dimensional black hole 
which is a solution of string theory. Let us concentrate on that solution 
here. As the $x^1$-coordinate runs from $-\infty$ to $+\infty$ it is to be
identified with the tortoise coordinate of the black hole. For details of
the solution see the next section. In the tortoise coordinate the metric
looks like ( we put $x^1\equiv\sigma$ in the next section)
\eq{345}{ds^2=(1-{2M\over r})(-dt^2+(dx^1)^2)}
where $(1-2M/r)=(1+{M\over\lambda}e^{-2\lambda x^1})^{-1}=\sqrt{-g}$. Also
$\grad^1\grad_1=(1+{M\over\lambda}e^{-2\lambda x^1})(\grad_1)^2$ and $\chi=
(Q\sqrt\pi/e)\,\theta\,(x^1+L/2)\,\theta\,(L/2-x^1)(1+{M\over\lambda}e^{-2
\lambda x^1})$. Now the task is to evaluate the integrals. For the time
being let us take $\varphi={\rm const.}$ which we deed in 
\cite{BGM} and try to calculate $V(L)$ for large $L$. Then the first
integral becomes clearly the same as the first one in (\ref{344a})
After a little thought the second one will also turned out to be the
same  as the second one in (\ref{344a}). To see this explicitly
just make a change of variable from $x^1$ to $x^1/L$. So in the limit
$L\to\infty$ the potential $V(L)$ doesn't really alter at all even
in this non-trivial background. It has been argued a long time ago that
finite curvature is like finite temperature and as the Schwinger model
doesn't show any non-trivial phase at finite temperature its behavior
is not expected to change in the presence of gravity. We proved here
this conjecture explicitly at least for a non-trivial background. 

We can now make some comments about the phases of the model. Since we
have a free space dependent parameter in our solution
we can see the nature of the spectrum and quark interactions
for various forms of this function. Again for simplicity let us put
the function $\varphi$ to be a constant. In that case the model
can be found only in two phases similar to the flat case, namely 
i) the constant $\to 0$: Then the mass of the boson is $m=e/\sqrt\pi$
and the potential $V(L)\sim{\rm const.}$ for large $L$. So this is
the familiar screened Coloumb phase and ii) the constant $\to\infty$:
Then the mass $m=0$ and also $V(L)=0, \forall L$. So the quarks become
essentially free and the mass-less boson can be interpreted as a
mass-less fermion \cite{Man}. Thus the fermions get liberated 
into the spectrum and this is an unconfining phase.

One interesting application of this model has already been discussed in 
\cite{GM}, i.e. in the problem of fermion - black hole scattering 
\cite{Strom}. There we have ignored the gravitational degrees of freedom 
and eventually the background was set to be flat \cite{PST,GM}. 
Now we shall study the same problem in an 
arbitrary curved background. For that we have to include the dilaton and 
the relevant action is essentially         
the string effective action in the sigma model metric when  
the dilaton kinetic term is dropped. 
\eq{346}{S\sim\intg\,[e^{-2\phi}(R+\lambda-{1\over 4}F^2)+i\overline\psi\slash D
     \, \psi]}
By a conformal rescaling of the metric $g_{\mu\nu}\to g_{\mu\nu}\exp(-2\phi)$ 
the action can be brought to a form similar to the Einstein-Maxwell action
with a modified space-time dependent coupling for the kinetic term 
of the electromagnetic field. The coupling involves the dilaton field $\phi$.
The metric equations would give rise to in general a curved background
and the problem really boils down to the study of Schwinger model in
such a background. Actually string theory offers many more copies
of the abelian gauge fields and the fermions can interact with all of
them. For simplicity we have restricted us to this case.

Now as above we first bosonize the fermionic fields and the resulting
action is exactly the same as (\ref{325}) with an additional kinetic 
term for the abelian gauge field $\sim-\exp\,(-2\phi)F^2$. So as is 
observed from the above general analysis that the essential spectrum
of the theory does not alter even after introducing an arbitrary curved
background the conclusions made in \cite{GM} still hold. So at least it
is clear that some quantum gravitational considerations are necessary to
address the question of whether the information is really lost.

\newsec{Temperature and entanglement entropy} 

There have been some attempts at calculating the entropy of quantum
fields in black hole backgrounds \cite{tHooft}, in contrast to the
more conventional Bekenstein entropy \cite{BEK}. The values thus obtained 
are contributions to the entropy of the black hole - field system.
These calculations have produced divergences \cite{UGLUM}. We shall 
see that similar phenomena occur in general for two dimensional black holes
also \cite{2d} in a different way. Simple 
Einstein action is trivial in two dimensions and related to the Euler number 
of the underlying manifold by the Gauss-Bonnet 
theorem. The situation is slightly non-trivial in string theory where we have 
an extra scalar field, the dilaton, coupled with the curvature. The model can   
be extended to have electromagnetic interactions and fermions. Actually
if we consider e.g. the heterotic string on eight torus the resulting effective
theory in two dimensions automatically has many copies of Abelian gauge fields
and also have fermions. Many black hole solutions of this model have been 
found with non-zero charge.

An eternal black hole can generally be taken as
\eq{347}{ds^2=-g_{tt}(r)\,dt^2+g_{rr}(r)\,dr^2}
together with a dilaton $\varphi$.
A classical black hole has a  horizon  beyond which  nothing
can leak out. This suggests that it can be assigned a zero
temperature. But the relation between the area
of the horizon and the mass and other parameters like the charge
indicates a close similarity \cite{BCH} with
the  thermodynamical laws, thus allowing the definition of a temperature.
This analogy was understood as being of quantum origin and made  quantitative
after  the  discovery of Hawking radiation \cite{HAWK1}. The associated
Hawking temperature vanishes only in the classical limit.
The thermodynamics of black holes has been extensively studied since then.

Most of the studies were first made for the simplest kind of black
hole, {\it viz.}, the Schwarzschild space-time.
Of more recent interest is the case of the so-called extremal black holes
which have peculiarities not always present in the corresponding
non-extremal cases \cite{PRES,GM1}.
For extremal Reissner - Nordstrom black holes, e.g.,
the na\"{\i}vely defined temperature is zero, but
the {\it area}, which is usually thought of as the entropy, is nonzero.
For extremal  dilatonic  black  holes,  where  the temperature is
{\it not} zero, the area vanishes. As is well known that under some
approximations two
dimensional black holes appear in the extremal limit of some dilatonic
black holes. 

In this section we shall  reexamine the temperature of a black hole 
\cite{TEM}. 
We discuss the conical singularity approach in detail.
The best known method of calculating the temperature of a black hole is
through the relation with surface gravity. To
distinguish this  temperature from those arising in other
approaches, we may call it the Hawking temperature. Here
\bb
T={1\over 2\pi\sqrt{g_{rr}}}\left.{d\sqrt{-g_{tt}}\over dr}\,
\right|_{r=r_h}\ee
where, $r=r_h$, describes the horizon.
However, there are other approaches to the temperature, and these
must be considered in view of the peculiarities of extremal black
holes.

First we consider the question of a conical singularity on passing to
imaginary time. The metric
\bb
ds^2= dr^2 + r^2d\,\theta^2,\label{pol}
\ee
which describes the flat Euclidean metric in polar variables, can be
supposed to describe distances on the surface of a cone. The cone has a
singularity at its tip $r=0$, except in the limiting case when the cone
opens out as a plane. In this situation $\theta$ has a periodicity $2\pi$,
so one may say that the conical singularity is avoided by making $\theta$ an
angular variable with this period. This is relevant for black holes because
such a singularity tends to arise in the Schwarzschild and in the
non-extremal cases. In this approach, one passes to imaginary time and
writes the metric as
\bb
ds^2&=&g_{tt}(r)\,dt^2+g_{rr}(r)\,dr^2\nonumber\\
&=&\Omega(\,\rho)(d\rho^2 + \rho^2d\tau^2),
\ee
where $\tau=\alpha t$ with the constant $\alpha$ so chosen as to make
the conformal factor $\Omega$
finite at the horizon. For consistency, one requires
\bb
\rho= e^{\,\alpha r_*},
\ee
where $r_*$ is defined by
\bb
dr_*=\sqrt{{g_{rr}\over g_{tt}}}\;dr.
\ee
which implies that $\rho$ vanishes at the horizon, i.e., as $r_*\to\infty$ 
Now 
\bb
\Omega={g_{tt}\over\alpha^2\rho^2}
\ee
can be made finite at the horizon by making $\rho^2$ vanish linearly as the 
horizon is approached {\it i.e.,} by choosing $\alpha$.
Now for the conical singularity to be avoided, one must have
a periodicity of $2\pi$ for $\tau$, {\it i.e.}, a periodicity for
$t$ given by $2\pi/\alpha$. This corresponds to a temperature
\bb
T={\alpha\over 2\pi}\label{T}
\ee
which is the standard  result. Thus for a general
black hole, what may be called the {\it Unruh} temperature
may/may not agree with the Hawking temperature. In four dimensions, in
fact, there are many such extremal cases, where these things are very 
different. Let us now look for the expression of entanglement entropy
of scalar field in the background of such a general black hole.

As argued in \cite{HAWK2} the partition function for the
system can be defined by the (Euclidean) Lagrangian path integral for the
gravitational action coupled with matter fields. The dominant
contribution will come from the classical solutions of the action.
We may approximate the Euclidean action by taking something like
\eq{348}{S_E[\,g,\varphi,A,\Phi]=S_1[\,g_{cl},\varphi_{cl},A_{cl}]+
         S_2[\,g_{cl},\Phi] +\cdots.}
where $\Phi$ is the scalar field to be considered in the background
of the dilatonic black hole and $A$ stands for the background 
electromagnetic field. Quantum fluctuations of the metric, the
electromagnetic field and the dilatonic field are neglected and these 
variables are frozen to their classical values. The partition function 
can then be taken as
\eq{349}{\Z=e^{-S_1[\,g_{cl},\varphi_{cl},A_{cl}]}\int\D\Phi\;e^{-S_2[\,g_{cl}, 
      \Phi]}.}
We come now to the contribution of the scalar field $\Phi$ to the partition 
function. To calculate this we employ the brick-wall boundary condition 
\cite{tHooft}. In this model the field is cut off just outside the horizon. 
Mathematically,
\eq{350}{\Phi(x)=0\qquad {\rm at}\;\ r=r_h+\epsilon}
where $\epsilon$ is a small, positive, quantity and signifies
an ultraviolet cut-off. Let us set also an infrared cut-off (anticipating
the result)
\eq{351}{\Phi(x)=0\qquad {\rm at}\;\ r=\Lambda}
with $\Lambda >>r_h$.
The  wave  equation  for  a scalar field in this space-time reads
\eq{352}{{1\over\sqrt{-g}}\,\partial_{\mu}(\sqrt{-g}g^{\mu\nu}\partial_{\nu}\Phi)
       -m^2\Phi=0.}
A solution of the form
\eq{353}{\Phi=e^{-iEt}f_{E}(r)}
satisfies the radial equation
\eq{354}{{1\over\sqrt{-g}}\,\partial_r(\sqrt{-g}g^{rr}\partial_rf_E)+k_r^2f_E=0.}
An $r$-dependent radial wave number can be introduced from this equation by
\eq{355}{k_r(r,E)=[\,g^{tt}E^2-m^2]^{\,1/2}.}
Only such values of $E$ are to be considered here that the  above
expression  is  nonnegative. The values are further restricted by
the semi-classical quantization condition
\eq{356}{n_r\pi=\int_{r_h+\epsilon}^{\Lambda}dr\sqrt{g_{rr}}\,k_r(r,E),}
where $n_r$ has to be a positive integer.

Accordingly, the free energy $F$ at inverse temperature $\beta$
is given by the formula
\eqn{357}{\beta F&=&\sum_{n_r}\;\ln\,(1-e^{-\beta E})\nn
       &\approx &\int dn_r\ln\,(1-e^{-\beta E})\nn
       &=&-\int d\,(\beta E)\,(e^{\beta E}-1)^{-1}n_r\nn
       &=&-{\beta\over\pi}\int_{r_h+\epsilon}^{\Lambda}dr\sqrt{g_{rr}}
           \int {dE\over e^{\beta E}-1}\sqrt{\,g^{tt}E^2-m^2}.}
Here the limits  of  integration  for  $E$ are such that the
arguments  of  the  square  roots  are   nonnegative. 
The $E$ integral can be evaluated only approximately 
where  the lower limit of the $E$ integral has been approximately
set equal to zero. If  the  proper  value  is  taken,  there  are
corrections  involving  $m^2\beta^2$  which will be ignored here.
The entanglement entropy can be obtained from the formula
\eq{358}{S=\beta^2\,{\partial F\over\partial\beta}.}

Let us now consider a particular two dimensional black hole. We consider
the low energy effective action of heterotic string on eight torus. In
that action if we set all the gauge fields and moduli fields to zero
it takes the following form
\eq{359}{{\cal S}\sim\intg\, e^{-2\varphi}\,[\,R+4(\grad\varphi)^2+4\lambda^2]}
where, $1/\lambda$ is a length scale which the action inherits from
higher dimensions. This action is known to have black hole solution
\cite{gautam}. In fact, this black hole is the limiting eternal black
hole solution of the more interesting dynamical ones, which were obtained
in \cite{CGHS}. However, here we shall be talking only about the eternal
solution
\eq{360}{ds^2=-e^{-2\zeta}\,dx\,dy.}
$x,y$ represent the Kruskal-like coordinates if the solution is compared
with the `mock Schwarzschild' metric. In that case the Schwarzschild-like
coordinates can be introduced via the transformations
\eqn{361}{\lambda x &=& e^{\,\lambda\sigma^+}\nn
       \lambda y &=& -e^{-\lambda\sigma^-}}
where, $\sigma^{\pm}=t\pm\sigma$, are the light cone coordinates. The
coordinate $\sigma$ is like the Schwarzs- child - radial coordinate. The
conformal factor in front of the metric is given by $\exp(\,2\zeta\,)=-
\lambda^2xy+M/\lambda$. The horizon of this black hole is at $y=0$, the
curvature singularity is at the space-like curve $xy=M/\lambda^3$ and
the asymptotic region is described by $\sigma=\infty$. $M$ represents
the mass of the black hole. To have a more familiar form let us make this
coordinate redefinition 
\eq{362}{1-2M/r=(1+Me^{-2\lambda\sigma}/\lambda)^{-1}.}
This puts the metric in the form
\eq{363}{ds^2=-(1-{2M\over r})\,dt^2+{1\over 4\lambda^2 r^2}\,(1-{2M\over r}
        )^{-1}dr^2}
The horizon is mapped to $r=2M$, the curvature singularity is at $r=0$
and the asymptotic regions are described by $r=\infty$. This form is
very suggestive to compare the solution with the `mock Schwarzschild'
metric. Now let us try to estimate its temperature and entanglement 
entropy following the general procedure described above. First of all
the Unruh temperature can be fixed easily. the variable $\rho$ 
introduced above is $\rho=\exp\,(\alpha\sigma)=(r-2M)^{\alpha/2M}$. So
$\Omega=(1-2M/r)/\alpha^2\rho^2$ is finite at the horizon provided
we set $\alpha=\lambda$. This in turn implies $T=\lambda/2\pi$. 
Fortunately, for this solution the Hawking temperature agrees with the
Unruh temperature and the conical singularity can be removed by this
choice of the angular periodicity. Using the general formula above
we can also calculate the entanglement entropy for this black hole.
The free energy is given by 
$$F\sim -{1\over\lambda\beta^4}\,\ln{\Lambda\over\epsilon}$$ 
where $\beta=2\pi/\lambda$. We neglected the other proportionality
factors in the expression of free energy. So the entropy
is $S\sim\lambda^2\ln(\Lambda/\epsilon)$, it receives divergences 
from the ultraviolet as well as from the infrared regions \cite{entrf}.

There is another way to fix the temperature of a black hole which is from
the study of Hawking radiation. For this we have to calculate the 
expectation value of the number operator of say, a scalar field in
the Schwarzchild-like coordinate system in the vacuum of the Kruskal-like
observer. The celebrated result is that it would come out to be a Bose
distribution which is interpreted as a thermal radiation coming out of
the black hole. The distribution function then corresponds to a definite 
temperature. As we have already mapped the black hole to a `Mock 
Schwarzchild'-like form (at least near the horizon $r\approx 2M$ there is 
hardly any difference) the space-time can also be mapped to a Rindler one.
Essentially, the Kruskal construction given above is nothing but the 
familiar transformations between the Rindler and the flat spaces near the
horizon. To see this explicitly let us scale the time $t\to t/\lambda$,
and define the Rindler coordinate $\eta=\sqrt{1/\lambda M}\exp(\lambda\sigma)$.
Then the line element near the horizon takes the standard form
\eq{364}{ds^2\approx -{\lambda\over M}\,dx\,dy=-\eta^2\,d\tau^2+d\eta^2}
where $\tau=\lambda t$.
As it is well known that the Rindler observer sees a thermal bath of temperature
$1/2\pi$ and we have scaled the time coordinate by $\lambda$ the actual
temperature which the asymptotic observer sees is $\lambda/2\pi$ consistent
with the earlier observations. 

Finally it is important to see that why the entanglement entropy is so
important in the study of black holes and its semi-classical properties.
We can very simply demonstrate that the entanglement between the different
partitions in the space-time leads us to the consideration of a thermal
picture. The Hawking radiation is an outcome of such a partitioning.
As we have seen that the space-time near to the horizon is essentially
flat and can be mapped to a Rindler one, the physics of that region is
captured by the transformations between these two frames. We shall be
calling these two observers as the `flat' and the `Rindler' observers
accordingly. As the flat observer does not have any coordinate singularity
at the horizon she has access to both sides of the horizon. However, the
Rindler observer doesn't have access across the horizon. So from the
point of view of the flat observer the Rindler space-time is only a
part of the entire space-time and the Rindler observer has access only
to that part. We shall see explicitly the effect of such a splitting.
If we take the flat space coordinate as $X,Y$ with $x=T+X$ and 
$Y=T-X$ then the Rindler mappings are
\eqn{365}{T&=&\sqrt{M\over\lambda}\ \eta\sinh\tau\nn
       X&=&\sqrt{M\over\lambda}\ \eta\cosh\tau.}
Then a small translation of the Rindler time $\delta\tau=\epsilon$
with $\delta\eta=0$ corresponds to the following flat space-time
transformations
\eqn{366}{\delta T&=&X\epsilon\nn
       \delta X&=&T\epsilon.}
It is in fact, the Lorentz boost along the space direction $X$. The
generator is 
\eq{367}{H_R=\int_{-\infty}^{+\infty}dX\,({\cal H}\/X-{\cal P}\/T)}
where ${\cal H,P}$ are respectively the Hamiltonian and momentum operators
in the flat space-time.
Now the generator being a conserved operator in time the integral can
be evaluated at any time slice. Let us set it at $T=0$. Then it takes
the simple looking form $H_R=\int dX X\,{\cal H}(X,0)$. Now if we look
only along the axis $T=0$ the space of the Rindler observer is split
into the two parts $X>0$ and $X<0$. So we split the generator also
accordingly as $H_R=H_>-H_<$, where
\eqn{368}{H_>&=&\int_{-\infty}^{+\infty}dX\ \theta\,(+X)\,X{\cal H}\nn
       H_<&=&-\int_{-\infty}^{+\infty}dX\ \theta\,(-X)\,X{\cal H}.}
Now given the mode expansions of a scalar field of mass $m$ in terms
of a set of creation and annihilation operators in the flat space-time
it is possible to find out another set through the generator $H_R$
can be expressed as $H_R=\int d\omega\,\omega\,a_\omega^\dagger a_\omega$.
In fact, given the free field expansion as
\eq{369}{\Phi(X,T)=\int_{-\infty}^{+\infty}{dk\over\sqrt{4\pi k_0}}\ [\,b_k
      \exp\,(ikX-ik_0T)+{\rm h.c.}]}
with $k_0^2=m^2+k^2$ and $[\,b_k,b_{k'}^\dagger]=\delta_{kk'}$ the 
construction of $a_\omega$ is given in \cite{thooft2}
\eq{370}{b_k=\int_{-\infty}^{+\infty}{d\omega\over\sqrt{2\pi k_0}}\ a_\omega 
      \exp\left(-i\omega\ln{k_0+k\over m}\right).}
This construction gives $[\,a_\omega,a_{\omega'}^\dagger]=\delta_{\omega
\omega'}$. Now the general problem of splitting is as follows: let the
annihilation operators corresponding to the Hamiltonians $H_>$ and $H_<$ 
are $a_>$ and $a_<$ respectively. Let us also impose the condition that
these two regions $>$ and $<$ are completely disjoint and there is no
correlation between them. So we have the following set of conditions
\eqn{371}{[\,a_>(\omega),a_<(\omega')\,]&=&0\nn\
       [\,a_>(\omega),a_<^\dagger(\omega')\,]&=&0\nn\
       [\,a_>(\omega),a_>^\dagger(\omega')\,]&=&\delta_{\omega\omega'}\nn\
       [\,a_<(\omega),a_<^\dagger(\omega')\,]&=&\delta_{\omega\omega'}\nn
       \int_{-\infty}^{+\infty}d\omega\ \omega\,a_\omega^\dagger a_\omega
       &=&\int_{-\infty}^{+\infty}d\omega\ \omega\left(\,a_>^\dagger(\omega) 
       a_>(\omega)-a_<^\dagger(\omega)a_<(\omega)\right).}
If now the operators $a_>$ and $a_<$ are related to the flat space
operators $a_\omega$ and $a_\omega^\dagger$ as follows
\eqn{372}{a_>(\omega)&=&\sum_{\omega'}~A_{\omega\omega'}a_{\omega'}
           +B_{\omega\omega'}a_{\omega'}^\dagger\nn
       a_<(\omega)&=&\sum_{\omega'}~C_{\omega\omega'}a_{\omega'}
           +D_{\omega\omega'}a_{\omega'}^\dagger}
we can try to solve for these coefficients from the above conditions.
Obviously the general solution can not be obtained, but we can look
for at least a special solution. A one parameter special solution is
\eqn{373}{A_{\omega\omega'}=\delta_{\omega\omega'}{1\over\sqrt{1-e^{-\beta
      \omega}}}&,&B_{\omega\omega'}=\delta_{\omega,-\omega'}{e^{-\beta
      \omega/2}\over\sqrt{1-e^{-\beta\omega}}}\nn
      C_{\omega\omega'}=\delta_{\omega,-\omega'}{1\over\sqrt{1-e^{-\beta
      \omega}}}&,&D_{\omega\omega'}=\delta_{\omega\omega'}{e^{-\beta
      \omega/2}\over\sqrt{1-e^{-\beta\omega}}}.}
This solution corresponds to Hawking radiation. To see this explicitly
let us define the vacuum for the flat observer as $b_k|0>=a_\omega|0>=0$.
In this vacuum one of the Rindler observer, say in the region $>$, would
see a spectrum as
\eq{374}{<0|\,a_>^\dagger(\omega)a_>(\omega')\,|0>\,=\sum_\lambda\,B_{\omega
      \lambda}^*B_{\omega'\lambda},}
which for the special solution boils down to the form a thermal spectrum
obeying Bose-statistics
\eq{375}{\sum_\lambda\,B_{\omega\lambda}^*B_{\omega'\lambda}=\delta_{\omega
      \omega'}\,{1\over e^{\beta\omega}-1}.}
Here the parameter $\beta$ should be identified with the temperature.
As from the other considerations the temperature is known to be $\lambda/
2\pi$ the Hawking spectrum is explicitly known in this case.

\newsec{Discussion}

We see in this paper that QED$_2$ in non-trivial backgrounds does not
show any sign of being in a different phase other than those in the
flat cases. It is still hard to give a completely general proof of this
in an arbitrary background but we could establish the fact at least in
some black hole background. Possibly explicit calculations of
Green functions could be possible in this case but it is still
hard to draw conclusions even from that as we have experienced already
in the flat case.
As a possible application of these solutions we considered the fermion-
black hole scattering problem. The above conclusion have an important
implication on this, however. We could draw conclusions that the problem
of information loss for fermions can not be solved if we treat gravity
classically and this motivates further the search for a viable quantum theory
of gravity to avoid such unpleasent events around these special
backgrounds.

We also did analysis about the nature of QED$_2$ in this special background.
As the spectrum of the abelian theory contains only a scalar field the
analysis is sufficiently simplified. We showed that why people are so
much interested about the entropy coming from the splitting of fields.
The essential reason is that the Hawking's calculations are also entirely
based on such splittings of space-times. The field theoretic calculations of entropy,
however, show divergences coming both from the ultraviolet and the infrared
regions. Many string theoretic calculations have been done afterwards and
they're potentially finite \cite{entrs}. However, it has to be remembered that
the calculations involving such splittings are associated with scales.
e.g. if the characteristic length scale of one of the regions gets too
small compared to that of the other, then essentially there is no splitting
and there should be hardly any thermal spectrum seen by the Rindler
observer. In that case we expect that the temperature should go to
infinity (to have a vanishing spectrum) as the region inside the horizon 
gradually shrinks to smaller
and smaller sizes by emitting Hawking radiations. Actually somewhere,
the field theoretic framework possibly breaks down and we need a more
microscopic theory such as string theory to probe. The divergences are
possible indications of that failure of field theory to handle the
high energy regimes if and only if it becomes impossible to accommodate
those divergences in some reasonable renormalization scheme, which is
also far from our sight essentially in higher dimensions.  Also the fact
that the temperature goes to infinity as we go towards the end point
of Hawking radiation indicates that somewhere in between the thermal
description should break down as well. We do not yet have a good picture
of either of these two problems \cite{future1} and hopefully in the coming 
years we will have a unifying point of view \cite{future2,future3} of 
resolving both these problems
of information loss and black hole thermodynamics simultaneously.

\vspace{.5cm}
\centerline{\large\bf Acknowledgment}\medskip

The author would like to acknowledge several useful and critical discussions
with various people which include Subir Mukhopadhyay, Amit Kundu,
Gautam Bhattacharya and Parthasarathi Mitra. The author is especially 
thankful to G.B. and P.M. for continuous encouragement and 
a careful reading of the manuscript and 
for giving various helpful suggestions and making criticisms. The author 
would also like to thank the International Centre for Theoretical Physics 
(ICTP), Trieste, for hospitality.

\newpage
%

%
%
\end{document}